\documentclass[conference]{IEEEtran}
\IEEEoverridecommandlockouts
\usepackage{cite}
\usepackage{amsmath,amssymb,amsfonts}
\usepackage{algorithmic}
\usepackage{graphicx}
\usepackage{textcomp}
\usepackage{xcolor}
\usepackage{subcaption}
\usepackage{enumitem}
\usepackage{multirow}
\usepackage[pagebackref=false,breaklinks=true,colorlinks,bookmarks=false]{hyperref}
\def\BibTeX{{\rm B\kern-.05em{\sc i\kern-.025em b}\kern-.08em
    T\kern-.1667em\lower.7ex\hbox{E}\kern-.125emX}}
\begin{document}

\title{Support Vector Machine for Person Classification Using the EEG Signals}

\author{\IEEEauthorblockN{1\textsuperscript{st} Naveenkumar Venkataswamy}
\IEEEauthorblockA{\textit{Electrical and Computer Engineering} \\
\textit{Clarkson University}\\
New York, USA \\
venkatng@clarkson.edu}
\and
\IEEEauthorblockN{2\textsuperscript{nd} Masudul H Imtiaz}
\IEEEauthorblockA{\textit{Electrical and Computer Engineering} \\
\textit{Clarkson University}\\
New York, USA \\
mimtiaz@clarkson.edu }
}

\IEEEoverridecommandlockouts
\IEEEpubid{\makebox[\columnwidth]{979-8-3503-3607-8/23/\$31.00 ~\copyright2023 IEEE \hfill}
\hspace{\columnsep}\makebox[\columnwidth]{ }}
\maketitle
\IEEEpubidadjcol

\begin{abstract}
User authentication is a pivotal element in security systems. Conventional methods including passwords, personal identification numbers, and identification tags are increasingly vulnerable to cyber-attacks. This paper suggests a paradigm shift towards biometric identification technology that leverages unique physiological or behavioral characteristics for user authenticity verification. Nevertheless, biometric solutions like fingerprints, iris patterns, facial and voice recognition are also susceptible to forgery and deception. We propose using Electroencephalogram (EEG) signals for individual identification to address this challenge. Derived from unique brain activities, these signals offer promising authentication potential and provide a novel means for liveness detection, thereby mitigating spoofing attacks. This study employs a public dataset initially compiled for fatigue analysis, featuring EEG data from 12 subjects recorded via an eight-channel OpenBCI helmet. This dataset extracts salient features from the EEG signals and trains a supervised multiclass Support Vector Machine classifier. Upon evaluation, the classifier model achieves a maximum accuracy of 92.9\%, leveraging ten features from each channel. Collectively, these findings highlight the viability of machine learning in implementing real-world, EEG-based biometric identification systems, thereby advancing user authentication technology.
\end{abstract}


\section{Introduction} With the rapid advancement of the internet, the security of personal information has emerged as an indispensable component within security systems. In recent times, a plethora of personal authentication methods have been introduced to bolster data security. However, traditional forms of personal identification such as passwords, Personal Identification Numbers (PINs), Identification tags (IDs), and signatures, have proven inadequately reliable in meeting security requisites due to their susceptibility to risks like leakage, theft, or forgery \cite{10.1147/sj.403.0769}. As a remedy, the adoption of biometric technology has surged; this technology leverages distinctive physiological or behavioral attributes of the human body to verify users and mitigate the vulnerabilities associated with traditional authentication methods.

The present landscape employs morphological biometrics such as facial features, fingerprints, voice patterns, and iris characteristics in authentication processes. A noteworthy effort in this realm was put forth by authors in \cite{jain1999multimodal}, who proposed a multimodal biometric system integrating fingerprints, facial recognition, and speech analysis to enhance resistance against attacks. It is essential to acknowledge, however, that these systems remain susceptible to spoofing attempts \cite{Chingovska2014BiometricsEU}.

Emerging on the horizon are novel biometric authentication forms- namely, Electrocardiogram (ECG), Electroencephalogram (EEG), and Electromyography (EMG). Distinguished by their foundation in living body signals, these methods share defining attributes: universality (ubiquitous across all humans), uniqueness (distinct for each individual), and circumvention (resistant to falsification). 

Among these, EEG harnesses the intricate electrical activity of the brain, signifying its cognitive functions. The EEG signals exhibit a high degree of randomness and potentially harbor valuable insights into brain states \cite{subha2010eeg}. Variants of EEG capture devices exist, varying in electrode configurations, with some utilizing dry electrodes and others employing wet electrodes. These devices gauge cerebral cortex electrical activity while a subject performs specific actions. Placed non-invasively on a participant's scalp, voltage-sensitive electrodes intercept individualized brain waves \cite{losonczi2014embedded}. These neural signals are underpinned by a person's unique neural pathway patterns, rendering them nearly impossible to replicate \cite{palaniappan2007biometrics}. Moreover, the signals are influenced by mood and mental disposition, rendering coercion and duress ineffective in their acquisition \cite{alsolamy2016emotion}. Additionally, these signals bear a genetic component, ensuring their distinctiveness and constancy over time. As a result, EEG signals have garnered substantial attention in the realm of biometric authentication, exhibiting advantages over conventional modalities such as fingerprints, irises, and facial features.

This paper presents a comprehensive framework for EEG-based user authentication utilizing Machine Learning techniques. The EEG data of 12 subjects performing an oddball auditory task is collected using wireless OpenBCI equipment. Given the inherent inter-individual variability in brain activity, diverse machine learning and deep learning models have been employed in previous studies \cite{jayarathne2017survey}. Notable examples encompass a multiclass support vector machine (SVM) approach yielding a classification accuracy of 94.44\% \cite{bashar2016human}; a Naive Bayes model facilitating decision-making in EEG-based person authentication \cite{he2009independent}; utilization of SVM classifiers with various kernels achieving up to 90.7\% accuracy via fuzzy entropy \cite{mu2017comparison}; a convolutional neural network (CNN) method attaining 97\% accuracy in EEG-based person authentication \cite{yu2019eeg}; and a fusion of steady-state visual evoked potential (SSVEP) and event-related potential (ERP) features utilizing long short-term memory (LSTM) networks resulting in a commendable 91.44\% verification accuracy \cite{puengdang2019eeg}.

This study proposes an SVM classifier technique for authentication tasks, employing a publicly available dataset \cite{0sjm-r993-21} originally collected for fatigue analysis. The proposed multi-class SVM, armed with 27 principal components and 12 labels (subject numbers), achieves a classification accuracy of 92.9\% by extracting crucial features from raw EEG signal data. The ensuing sections delineate the experimental setup and design in Section 2.1, followed by a presentation of signal processing, feature extraction, and model evaluation methodologies in Section 2.2.

\section{Materials and Methods}
\subsection{Datasets} 
The research presented in this paper utilizes a publicly accessible dataset initially compiled by \cite{0sjm-r993-21}, which was originally designed to assess individual mental fatigue \cite{ramirez2021evaluation}. Named 'EEG AND EMPATICA E4 SIGNALS FIVE MINUTE P300 TEST AND FAS SCOREs,' this dataset contains information from 12 healthy participants. The subjects in this dataset have an average age of 22 years, with a standard deviation of ±3 years, reflecting a young adult demographic. This careful data selection ensures a homogeneous sample set, reducing the potential for confounding variables associated with age-related cognitive differences. Consequently, the dataset serves as a robust foundation for our analysis, enabling us to conduct a focused investigation into the utilization of EEG signals for user authentication.

\subsubsection{Experimental Design}
Twelve participants, with an average age of 22 years and a standard deviation of ±3 years, took part in the study. These participants were provided with a consent form that explained the data collection procedure and informed them about their rights. After submitting the consent form, the participants were then asked to complete a questionnaire and undergo tests on the Fatigue Rating Scale (FAS). Subjects were instructed to maintain a comfortable sitting position in a relaxed environment, consistent with the workplace setting (e.g., a classroom or office). During this time, their electroencephalographic (EEG) signals were measured by placing an Ultracortex “Mark IV” EEG headset on their heads.

\begin{figure}[h]
\centering
    \includegraphics[width=8.5cm]{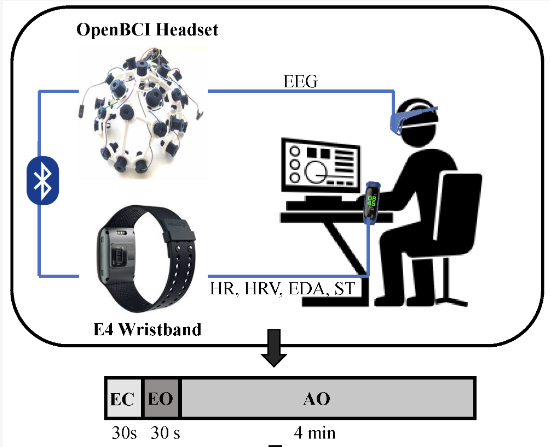}
    \caption{\footnotesize The design concept for EEG signal acquisition experiments. Biometric data were collected using OpenBCI in the relaxed state (E.C. and E.O.).) and the A.O. task. In the A.O. tasks, frequent and infrequent stimuli are presented randomly in an 80:20 ratio. The data is transferred to the P.C. via Bluetooth with the OpenViBE software \cite{ramirez2021evaluation}.}
    \label{fig:setup}
    \vspace{-3mm}
 \end{figure}

The participants underwent a 5-minute recording session, which consisted of 30 seconds with their eyes closed (E.C.), followed by 30 seconds with their eyes open (E.O.), and then a 4-minute auditory task (A.O.) designed to elicit P300 waves. To distinguish between the E.C., E.O., and A.O. tasks, a low-pitched tone was played for three seconds to mark the transitions. The experimental setup is visually represented in Fig~\ref{fig:setup}.

In total, a 120-second auditory stimulus, which consisted of 2 different tones (24 non-frequent and 96 frequent) with a probability proportion of 80:20, as suggested in \cite{ito1996skin, debener2005novel}. These stimuli were presented for one second, with an inter-stimulus period of one second. Volunteers were asked to use headphones during recordings to avoid distraction from ambient noise. Each participant completed one trial of the protocols explained. A stimulus presentation application was designed using OpenViBE software to enable the real-time acquisition, filtering, processing, classification, and visualization of brain signals \cite{renard2010openvibe}.

\subsubsection{EEG Signal Acquisition}
\begin{figure}[h]
\centering
    \includegraphics[width=8.5cm]{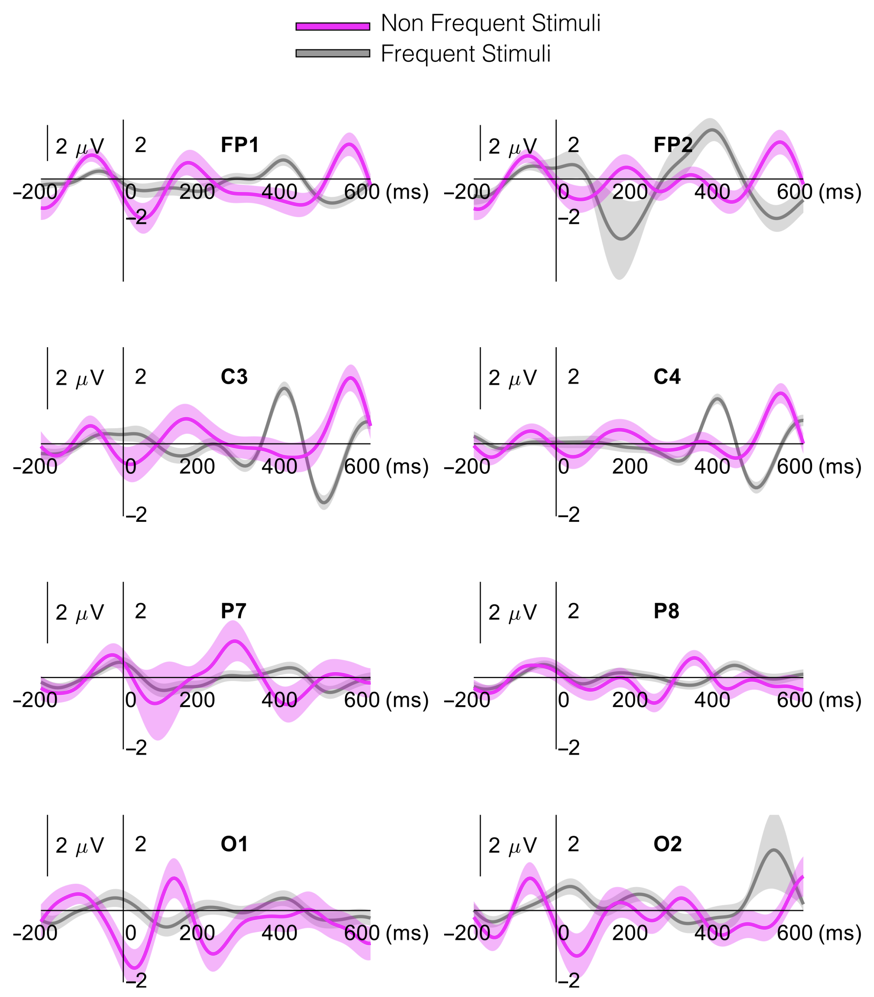}
    \caption{\footnotesize Grand average representation of the P300 wave across all the participants. Eight traces (one per channel) are presented for the frontal(F), center(C), parietal (P), and occipital(O) electrodes of the left and right hemispheres of the brain. The shaded area represents standard error across traces \cite{ramirez2021evaluation}}
    \label{fig:signalproces}
    \vspace{-3mm}
 \end{figure}

EEG signals were acquired using OpenBCI wireless hardware and a Cyton board, allowing the system to capture dry electrode EEG signals at a sampling rate of 250 Hz. To position the electrodes, we utilized an Ultracortex Mark IV headset, known for its compatibility with any OpenBCI board and its ability to record EEG signals in various configurations \cite{aldridge2019accessible}. These configurations include the standard 10-20 system's 8 EEG channels: FP2, FP1, C4, C3, P8, P7, O1, and O2, along with two attachable reference electrodes—one in each earlobe. Subsequently, these signals underwent filtering, which involved a 60 Hz notch filter to eliminate powerline noise and a fourth-order Butterworth bandpass filter ranging from 0.1 to 100 Hz. Fig~\ref{fig:signalproces} illustrates the p300 waveform, showcasing the average response across all subjects to both frequent and non-frequent stimuli during the A.O. trial, recorded at eight channels using OpenViBE software.

\subsection{Methods}
The study methodology employed in this article follows a structured, stepwise approach: beginning with signal filtering to minimize noise and extraneous frequencies, segmenting EEG signals in intervals of time, extracting key features from these segments, choosing the most informative attributes, and constructing a classifier to identify users.
\subsubsection{Filtering}
The EEG signals underwent pre-processing before analysis, aiming to reduce unwanted noise and remove artifacts. To achieve this, OpenViBE was employed to filter the signals using a 60 Hz Notch filter to eliminate powerline noise and a fourth-order Butterworth bandpass filter ranging from 0.1 to 100 Hz. Following these initial filtering steps, all the signals underwent further cleaning using MATLAB's EEGLAB toolbox \cite{delorme2004eeglab} artifact subspace reconstruction (ASR) algorithm, with a parameter k=15 utilized to reduce the impact of significant artifacts effectively.

\subsubsection{Windowing}
After the EEG signal is filtered, a windowing process takes place. First, the EEG signal data is segmented into an overlapping square window, with each window containing 0.8 seconds of EEG data with a 0.4-second overlap with future and previous windows.
\subsubsection{Feature Extraction}
Feature extraction is an important stage before the classification stage. 7-time domain and 3-frequency domain features were calculated for each channel’s window, combining 80 features, including eight data channels. Each of the 10 features was selected based on the literature \cite{rahman2021multimodal,la2014stable}. The ten features are Root Mean Square, Standard Deviation, Skewness, Kurtosis, Hjorth activity, Hjorth complexity, Hjorth mobility \cite{cecchinseizure}, Shannon’s entropy, Spectral entropy \cite{zhang2008feature}, and Power Spectral Density \cite{demuru2020comparison}. 
\subsubsection{Dimensionality Reduction}
\begin{figure}[h]
\centering
    \includegraphics[width=8.5cm]{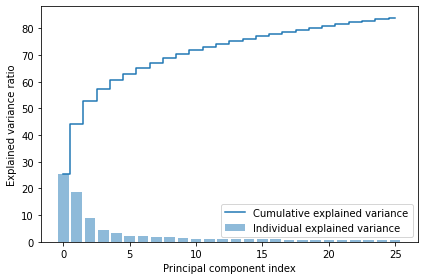}
    \caption{\footnotesize Variance plot from PCA. The X-axis shows the percentage variance explained by every 27 principal components. The Y-axis shows the cumulative sum.}
    \label{fig:PCA}
    \vspace{-3mm}
 \end{figure}

As we have a total of 80 features combined from 8 channels, there is a high probability of over-fitting \cite{roelofs2019meta} the model. Therefore, feature reduction is necessary to avoid over-fitting and eliminate irrelevant features. To standardize the input features and dimensionality reduction, we used principal component analysis (PCA) with 80\%, 85\%, 90\%, and 95\% explained variance. 95\% variance provided the best accuracy resulting in 27 principal components. Fig~\ref{fig:PCA} shows the variance contribution of each principal component.

\subsubsection{Classifier Implementation}
In our study, we proposed to use the 27 principal component features provided by PCA. We used the SVM classifier for the classification. Hence, we divided each individual’s samples into two parts: 80\% dedicated to learning and 20\% for testing.
\section{Results and Discussion}
For the learning phase, which is the next stage after the preprocessing and feature extraction phase, a model that can predict the class of the person is constructed using SVM, which is a discriminating learning technique that has given good performance in different biometric applications and therefore, for the classification task, we have applied the different SVM techniques (with the different kernel functions such as linear, polynomial and RBF) to compare with the different kernels’ classification accuracies. Each Kernal function has been modified and its parameters have been adapted to the obtained principal components. In the linear SVM, the regularization parameter c has been varied, which gives the best identification rate c=10 and therefore, the identification accuracy rate is equal to 81.1\%. The best identification rate is seen when c=1  for the polynomial kernel function. The Parameter degree is varied, keeping the parameter c of the polynomial function fixed to 1. Once the value of c and degree is fixed, the gamma parameter is varied, which gives an identification rate equal to 84.7\%. For best results for RBF kernel, the parameters c is set to 100 and gamma=0.01 and a level equal to 92.9\% is obtained. Table~\ref{table:SVMresults} summarizes the best recognition rates obtained with
the different kernels of SVM.

\begin{table}[!ht]
\scriptsize
\centering
\caption{\textbf{The best classification rates obtained with different SVM kernels}}
\label{table:SVMresults}
\begin{tabular}{|c|c|c|}
\hline
 \textbf{Linaer SVM} &  \textbf{Polynomial SVM} &  \textbf{RBF SVM} \\
 
  \hline
 81.1\% & 84.7\% & 92.9\% \\
 \hline

\end{tabular}
\vspace{-3mm}
\end{table}

This work proposes a machine learning framework for person identification using the support vector machine. By using the publicly available dataset, which is not collected with the intention of person identification, we can achieve a reasonable identification rate. We started by preprocessing the raw EEG signal and extracting the meaningful features suitable for identification purposes; then, an SVM was used to identify the user class. Although we couldn’t achieve the best accuracy with the dataset, it is observed that if the dataset is intently collected for user authentication purposes, we will be able to achieve much better results in terms of identification rate.
\section {Conclusions}
This paper proposes a machine-learning framework for human identification using a support vector machine. An experiment with publicly available 8-channel EEG data from 12 individuals showed the potential of the proposed method. The EEG data has been used to extract the features and later to identify the user. Based on the authentication features in both time and frequency domains, our work implemented fine-grained authentication efficiently. Furthermore, we systematically evaluated each parameter of SVM to implicit the user authentication. Based on the parameter choices, our evaluations demonstrate the advantages of combining EEG signals and machine learning to enhance authentication accuracy. The proposed study achieved authentication accuracy of up to 92.9\%. Although using a dataset not intended to perform user authentication showed good potential, the best authentication results can be achieved if the dataset is explicitly collected for authentication purposes.

\bibliographystyle{IEEEtran}
\bibliography{bibliography}

\end{document}